\documentclass[fleqn,usenatbib]{mnras}

\usepackage{newtxtext,newtxmath}
\usepackage[T1]{fontenc}

\DeclareRobustCommand{\VAN}[3]{#2}
\let\VANthebibliography\thebibliography
\def\thebibliography{\DeclareRobustCommand{\VAN}[3]{##3}\VANthebibliography}

\usepackage{graphicx}	% Including figure files
\usepackage{amsmath}	% Advanced maths commands
\usepackage{float}

\title[Characterization of acetonitrile ice II]{Characterization of acetonitrile ice irradiated by X-rays employing the PROCODA code: II. Desorption processes}

\author[Carvalho, Pilling and Gerasimenko]{
G. A. Carvalho,$^{1,2}$\thanks{E-mail: gacarvalho@utfpr.edu.br}
S. Pilling,$^{3}$
and S. Gerasimenko$^{1}$
\\
$^{1}$Departamento de Física, Universidade Tecnológica Federal do Paraná, Medianeira, 85884-000 PR, Brazil\\
$^{2}$ Programa de P\'os-Gradua\c{c}\~ao em F\'isica e Astronomia, Universidade Tecnol\'ogica Federal do Paran\'a, Jardim das Americas, 82590-300, Curitiba, PR, Brazil\\
$^{3}$Instituto de Pesquisa e Desenvolvimento, Universidade do Vale do Paraíba, São José dos Campos, 12244-000, SP, Brazil\\
}

\date{Accepted XXX. Received YYY; in original form ZZZ}

\pubyear{2015}

\begin{document}
\label{firstpage}
\pagerange{\pageref{firstpage}--\pageref{lastpage}}
\maketitle

\begin{abstract}
In this work, we focus on the study of radiation induced desorption processes that occurred in acetonitrile ice irradiated by broadband X-rays (6 eV to 2 keV) monitored by FTIR spectroscopy at different radiation fluences. In a previous work, we used the PROCODA code to derive the chemical evolution of the ice. Here, we have obtained that the acetonitrile desorbed column density is at least two orders of magnitude larger than the desorbed column densities of daughter or granddaughter molecular species at chemical equilibrium stage. This indicates that total desorption column density is mainly governed by the father molecule, as also previously hypothesized in experimental studies. This occurs basically because the acetonitrile column density is larger than the other ones. In particular, at chemical equilibrium acetonitrile desorption column density represents almost 98\% of the total, while it is close to 1\% for H, CN and CH$_2$, the species with larger molecular desorption percentages at chemical equilibrium. Another derived quantity is what we called intrinsic desorption rate, which is a number per second for individual species. Some of the larger intrinsic desorption rates were: CH$_3$CN ($6.2\times 10^{-6}$), CN ($6.2\times 10^{-6}$), H ($5.7\times 10^{-6}$), CH$_2$ ($5.7\times 10^{-6}$) and C$_2$N$_2$ ($4.4\times 10^{-6}$). These results help to put constrain in astrochemical models and can be also useful to clarify some astronomical radio observations.
\end{abstract}

\begin{keywords}
Physical and data processes: astrochemistry -- Physical and data processes: molecular processes -- X-ray: stars -- Software: simulations
\end{keywords}

%%%%%%%%%%%%%%%%%%%%%%%%%%%%%%%%%%%%%%%%%%%%%%%%%%

%%%%%%%%%%%%%%%%% BODY OF PAPER %%%%%%%%%%%%%%%%%%

\section{Introduction}

The radiation chemistry, important for the chemical evolution of molecular species in several astrophysical scenarios, involves a lot of phenomena, such as photodissociation, ionization and desorption to gas phase of astrophysical ices \cite{Moore1991Sep,Moore1996Sep,Kanda1999May,Hudson2006Jun,Pilling2010Nov,Vasconcelos2017Nov}. The last one is determining for molecular abundances at gas phase, e.g., the gas phase abundance of acetonitrile in protoplanetary disks is not entirely explained by pure gas phase reactions, so it is probably connected to the acetonitrile desorption on the surface of ice mantles \cite{Oberg2015Apr,Loomis2018Jun,Basalgete2021Dec,Basalgete2023Aug}. 

In fact, most molecular species are expected to be formed on the surface of the interstellar dust grains being later desorbed to gas phase. Therefore, the material in ice phase is a source for increasing molecular complexity of the gas phase in the interstellar medium. The James Webb Space Telescope (JWST) observations of interstellar ices in star-forming regions are revealing new details of ice evolution based on infrared (IR) features of frozen molecules. These observations in connection with laboratory investigations will give even further details on the acetonitrile gas phase formation.

Acetonitrile is a precursor of more complex nitrile molecules in space environments \cite{Gerakines2004Jul}. It has been detected in many different situations: in comets \cite{Huebner1974Dec,Goesmann2015Jul,Crovisier1998,Woodney2002May}, protoplanetary disks \cite{Oberg2015Apr,Bergner2018Apr}, interstellar medium \cite{Solomon1971Sep,Snyder1971Jan,Beltran2006Sep}, molecular clouds \cite{Willis2020Apr}, moons' atmosphere \cite{Lara1996Oct,Marten2002Aug,Thelen2019Feb,Thelen2020Nov,Iino2020Feb}, hot cores \cite{Bogelund2019Apr} and high/low mass protostars \cite{Taniguchi2020Jul,Andron2018Dec}. In addition, it is also among the five interstellar complex molecules desorbed by the frozen mantle of the circumstellar disk around the V883 Orionis protostar \cite{Lee2019Apr} and it is a commonly detected organic molecule in disks \cite{LeGal2019Nov}. 

Using the ALMA (Atacama Large Millimeter Array) radio telescope, organic molecules were tracked in the protoplanetary disks of five stars IM Wolf, AS 209, GM Aurigae, HD 163296 and MWC 480 \cite{Thelen2019Feb,Iino2020Feb,Willis2020Apr,Oberg2021Nov}. Authors have been able to trace simple organic molecules, including HCN, C$_2$H, H$_2$CO, HC$_3$N and also CH$_3$CN. The molecules have been found in the inner regions of the disks, where planets are likely to form. These molecules can, under the right conditions, serve as ``building blocks'' for the substances of terrestrial biochemistry: sugars, amino acids, and even components of ribonucleic acid (RNA).

Recently, vacuum ultraviolet (VUV) with energy in the range $7-13.6$ eV was used to derive the photodesorption yields of pure CH$_3$CN and CO and H$_2$O admixed acetonitrile \cite{Basalgete2021Dec}. Their results show that the photodesorption spectra of acetonitrile and its fragments (e.g., HCN, CN, CH$_3$) do not depend on the coadsorption of CH$_3$CN with CO and H$_2$O, which are dominant in astrophysical scenarios. More recently, the photodesorption of X-ray irradiated acetonitrile was also obtained \cite{Basalgete2023Aug}, their values help to understand the gas phase abundances of acetonitrile in protoplanetary disks. The thermal desorption of acetonitrile was also discussed in \cite{Corazzi2021Jun,Carvalho2022Feb}, where results indicate that thermal desorption depends weakly on ice composition, and it peaks around $120-300$ K.

In a previous work (hereafter paper I), we used the PROCODA code - a code developed to model the chemical evolution of the ice samples based on experimental data of irradiated astrophysical ice analogs - to study the X-ray irradiation of a pure acetonitrile ice sample \cite{Carvalho2022Sep}. To use the PROCODA code, we first need to fix a set of chemical reactions. In paper I, we consider 33 molecular species, between them 5 were observed by infrared techniques (4 irradiation products and acetonitrile) along with 240 chemical reactions (including dissociation, bimolecular and termolecular rates) in addition to 33 individual desorption process. 

In paper I, we detailed the molecular column densities, chemical equilibrium stage and effective reaction rates \cite{Carvalho2022Sep}. Here, we will focus on the desorption processes induced during ice irradiation. The experimental procedure was detailed before in several works such as \cite{Pilling2022Feb,Carvalho2022Sep,Pilling2023Jul,Pilling2023Jun,Pilling2023Aug}. The desorption processes addressed here together with experimental studies and astronomical observations can help us to understand the gas phase molecular abundances of acetonitrile in interstellar medium.

This work is organized as follows: in the next section, we give more experimental and model details, in section \ref{sec3} we present the results and in section \ref{sec4} we discuss some astrophysical implications of this work, finally in section \ref{sec5} we highlight our conclusions.

\section{Methodology}\label{sec2}

\subsection{Experimental data and the chemical reaction network}

The experiment was performed in a high-vacuum chamber coupled to the spherical grating monochromator (SGM) beamline at the facilities of the Brazilian Synchrotron Light Laboratory (LNLS). The beamline had a VUV+X-ray broadband spectrum (from 6 eV up to 2 keV) in order to simulate the processing of acetonitrile ice by soft X-rays in astrophysical environments. The total irradiation time was around $1.5\times 10^4$ s (final fluence of $1.5\times 10^{18}$ photons cm$^{-2}$), more experimental details can be consulted in \cite{Carvalho2020Oct}.

To use the PROCODA code to study irradiation of acetonitrile ice analogs by X-rays, a chemical reaction network must be provided. In paper I, we gave the entire set of 240 chemical reactions that are considered \cite{Carvalho2022Sep} and the effective reaction rates obtained. To build this reaction network, we analyzed several studies on acetonitrile irradiation and its photo-products \cite{Abdulgalil2013Jul,Hudson2004Dec,Hudson2001Apr,Hudson2008Oct,McElroy2013Feb,Bulak2021Mar,Carvalho2020Oct,Carvalho2022Feb}. All such reactions were considered to happen between neutral species. The chemical reaction network had 33 molecular species and included 240 chemical reactions. Besides the 240 chemical reactions, a desorption reaction for each species was also considered.

\subsection{Computational details}

Column density data were available only for five species: CH$_3$CN, CH$_3$NC, H$_2$CCNH, CH$_4$ and HCN. Experimentally, the summed desorption yield has been determined to be 0.29 molecules per photon, so the desorption provides one of the constraints that are imposed in the code calculations. The code applies other constraints besides the desorption yield, which is namely, minimization of $\chi^2$ for the column density data, mass conservation and chemical equilibrium stage. Those constraints are employed in a minimization process of a function; named score function (SF); that evaluates how close the system is to its minimum global solution. The score function is defined as follows,
\begin{eqnarray}
    &SF& = p_1\times \sum \frac{({\rm oCH}_3{\rm CN}_{\rm data} - {\rm oCH}_3{\rm CN}_{\rm model})^2}{{\rm oCH}_3{\rm CN}_{\rm data}} \nonumber \\
    &+& p_2\times \sum \frac{({\rm oCH}_3{\rm NC}_{\rm data} - {\rm oCH}_3{\rm NC}_{\rm model})^2}{{\rm oCH}_3{\rm NC}_{\rm data}} \nonumber \\
    &+& p_3\times \sum \frac{({\rm oH}_2{\rm CCNH}_{\rm data} - {\rm oH}_2{\rm CCNH}_{\rm model})^2}{{\rm oCH}_3{\rm CN}_{\rm data}} \nonumber \\
    &+& p_4\times \sum \frac{({\rm oHCN}_{\rm data} - {\rm oHCN}_{\rm model})^2}{{\rm oHCN}_{\rm data}} \nonumber \\
    &+& p_5\times \sum \frac{({\rm oCH}_{\rm 4data} - {\rm oCH}_{\rm 4model})^2}{{\rm oCH}_{\rm 4data}} \nonumber \\
    &+& p_6\times \left[(1-{\rm MSC}_f) + (1-{\rm MSCo}_f)  + (1-{\rm MSCo}_m)\right] \nonumber \\
    \label{SF}&+& p_7\times (1-{\rm DSC}) + p_8\times (1-{\rm SSC}),
\end{eqnarray}
where MSC$_f$ is the model mass similarity criterion, MSCo$_f$ and MSCo$_m$ are the column mass similarity criteria. The parameter DSC in equation \eqref{SF} is the desorption similarity criterion, obtained by comparing the experimental desorption yield \cite{Carvalho2020Oct} with the total molecular desorption yield computed by the model. Additionally, the parameter SSC is the slope similarity criterion. The first five terms in equation \eqref{SF} define the summed $\chi^2$ function for the data. The dimensionless parameters $p_1$ to $p_8$ employed in equation \eqref{SF} are the weights of each term.

More details on the entire procedure can be found in \cite{Pilling2022Feb,Pilling2023Jun,Pilling2023Jul}, including all steps of the code, such as, input data management, assessment of initial conditions for rate constants and search for a minimal global solution. It is worth to cite that a typical reaction in the chemical coupled system is written as 
\begin{eqnarray}
    \frac{dN_i}{dt} &=& \left[-{\rm DES}_i(t)  -\sum_{d1} k_{d1} N_i(t) -\sum_{d2} k_{d2}N_i(t) N_a(t)/L\right. \nonumber \\ &-&\left.\sum_{d3}k_{d3} N_i(t) N_a(t) N_b(t)/L^2 \right] \nonumber \\ &+& \left[  \sum_{p1} k_{p1} N_i(t) +\sum_{p2} k_{p2}N_i(t) N_a(t)/L\right. \nonumber \\ &+& \left.\sum_{p3}k_{p3} N_i(t) N_a(t) N_b(t)/L^2 \right],
\end{eqnarray}
where $dN_i/dt$ is the rate of change in the molecular column density of a given molecular species $i$, $N_i$ is the molecular column density for the $i$ species, and the $k$ values are the reaction rates for dissociation, bimolecular and termolecular processes ($p$ indices stand for production and $d$ for destruction processes), and $L$ indicates the sample average thickness in units of cm. The term ${\rm DES}_i$ is the column density desorbed to gas phase of the $i$-th species, which is modeled in the following way,
\begin{equation}\label{des_eq}
    {\rm DES}_i = k_{\textrm{DES},i} \Omega_i(t) N_i(t)
\end{equation}
where $k_{\textrm{DES},i}$ is called intrinsic desorption rate for the $i$-th molecular species given in Hertz units and $\Omega_i$ is the dimensionless surface coverage of the $i$-th species, defined as the column density of a given species over the sum of all column densities, $\Omega_i=N_i/\sum_i N_i$, and assumes values between 0 and 1. Behind equation \eqref{des_eq} is a working hypothesis that the sample is uniform, so the desorption column density of a given species is proportional to the column density of the same species. The desorption reactions were considered to be first-order type ones. To determine the total column density of desorbed species, we simply summed all individual desorption column densities. We calculate the total desorption yield by using the total desorption column density, sample area and radiation fluency.

\begin{figure*}
  \centering
  \includegraphics[width=0.9\linewidth]{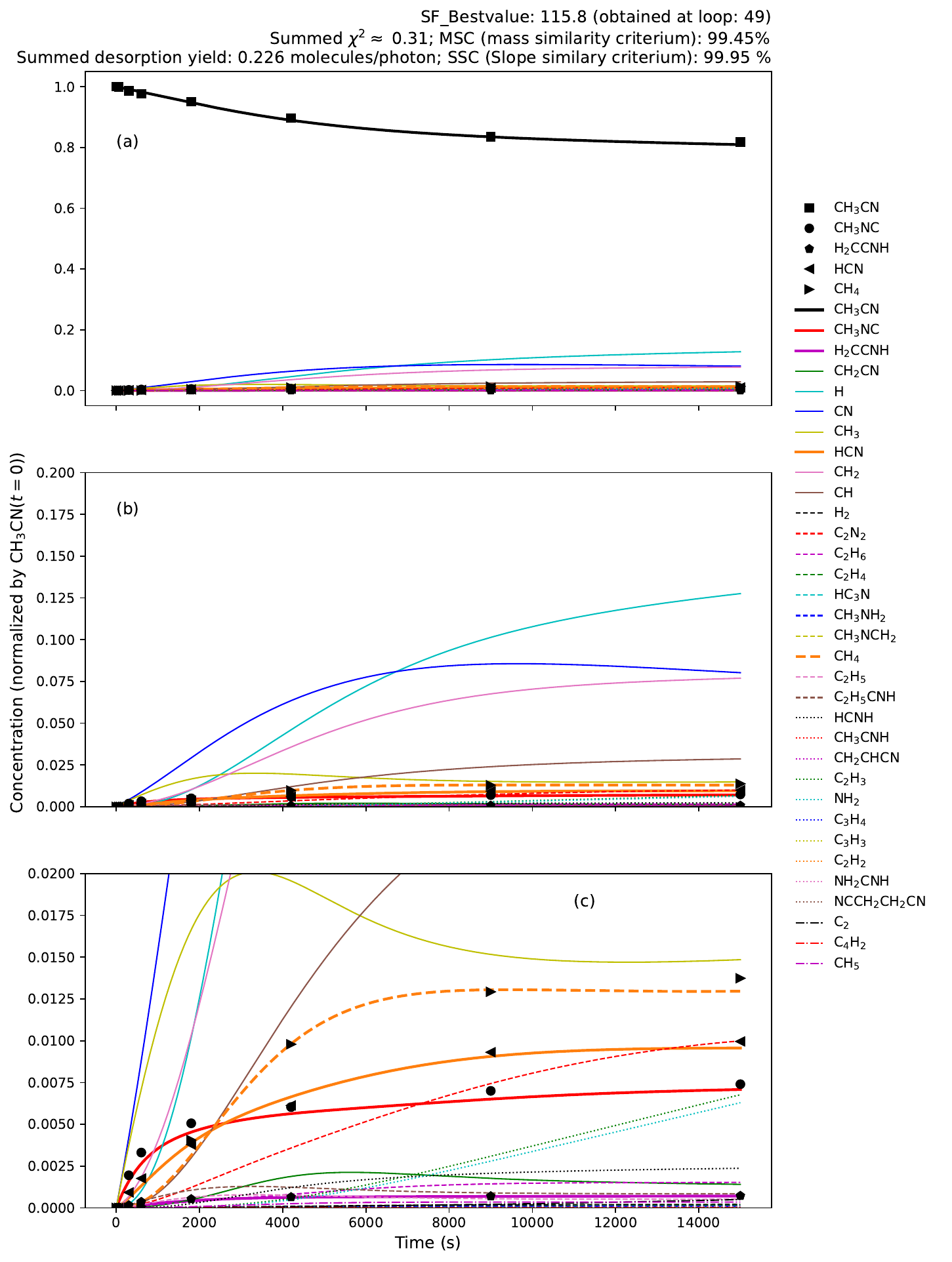}
    \caption{\label{concentration_zooms} Concentration as a function of time. In panel a) is visible that the concentration of the father molecule is higher compared to the other species. Panels b) and c) highlight regions where concentration scales are smaller. Concentration of all species are normalized by the initial concentration of CH$_3$CN.}
\end{figure*}

\begin{figure*}
  \centering
  \includegraphics[width=1.0\linewidth]{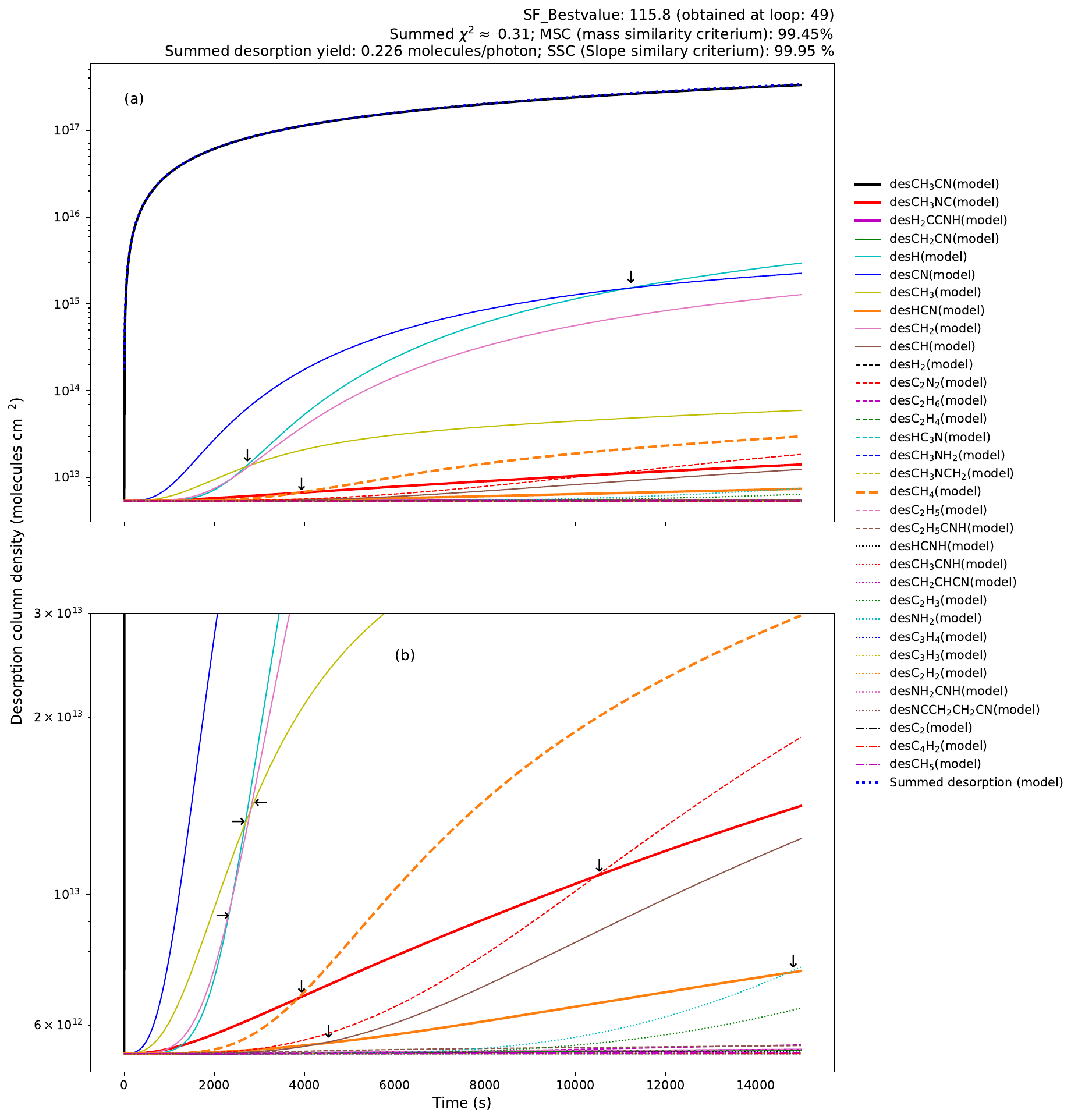}
    \caption{\label{desorption_density} Panel a) Molecular desorption induced by X-rays obtained by the best-fit model in the mapping of chemical evolution of acetonitrile irradiated ice. Panel b) Highlight of a region with smaller desorption column density values. In both panels, arrows indicate points where the desorption of a given species surpasses that of another one.}
\end{figure*}

\begin{figure*}
  \centering
  \includegraphics[width=0.49\linewidth]{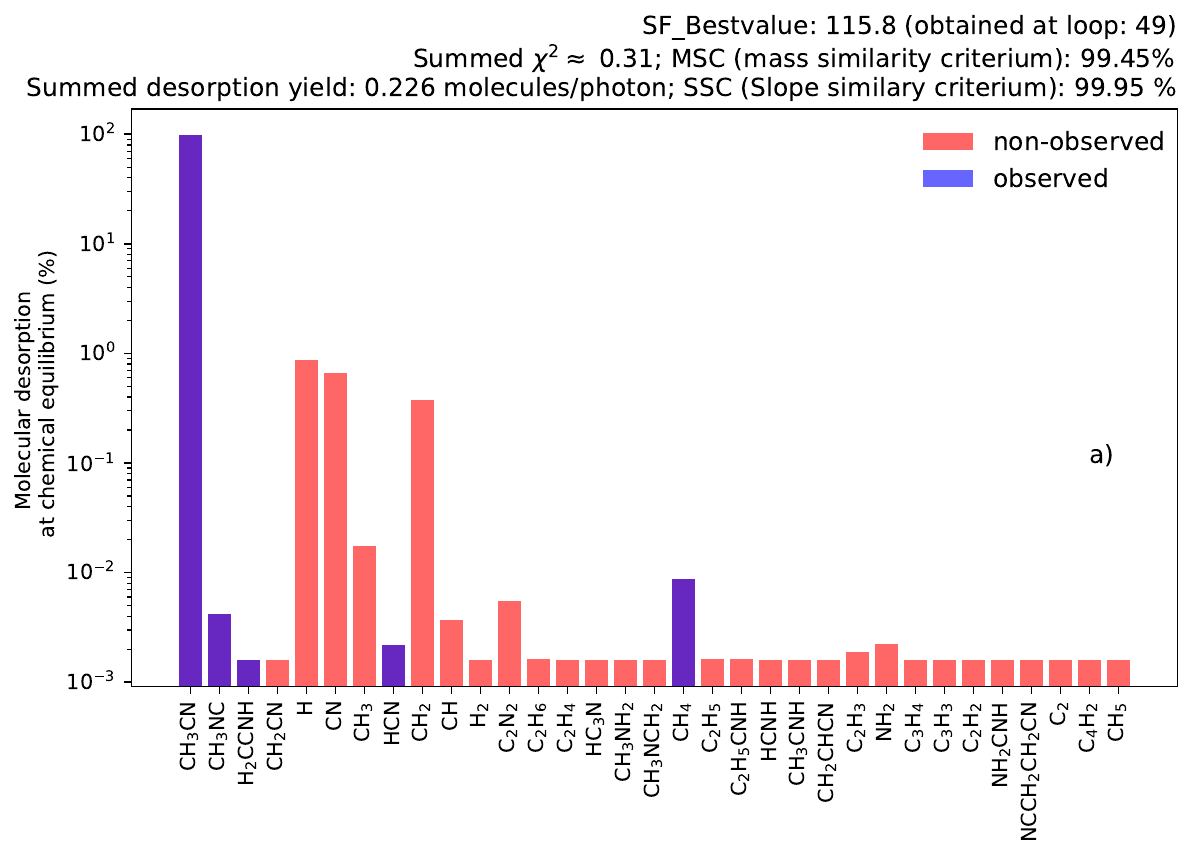}
  \includegraphics[width=0.49\linewidth]{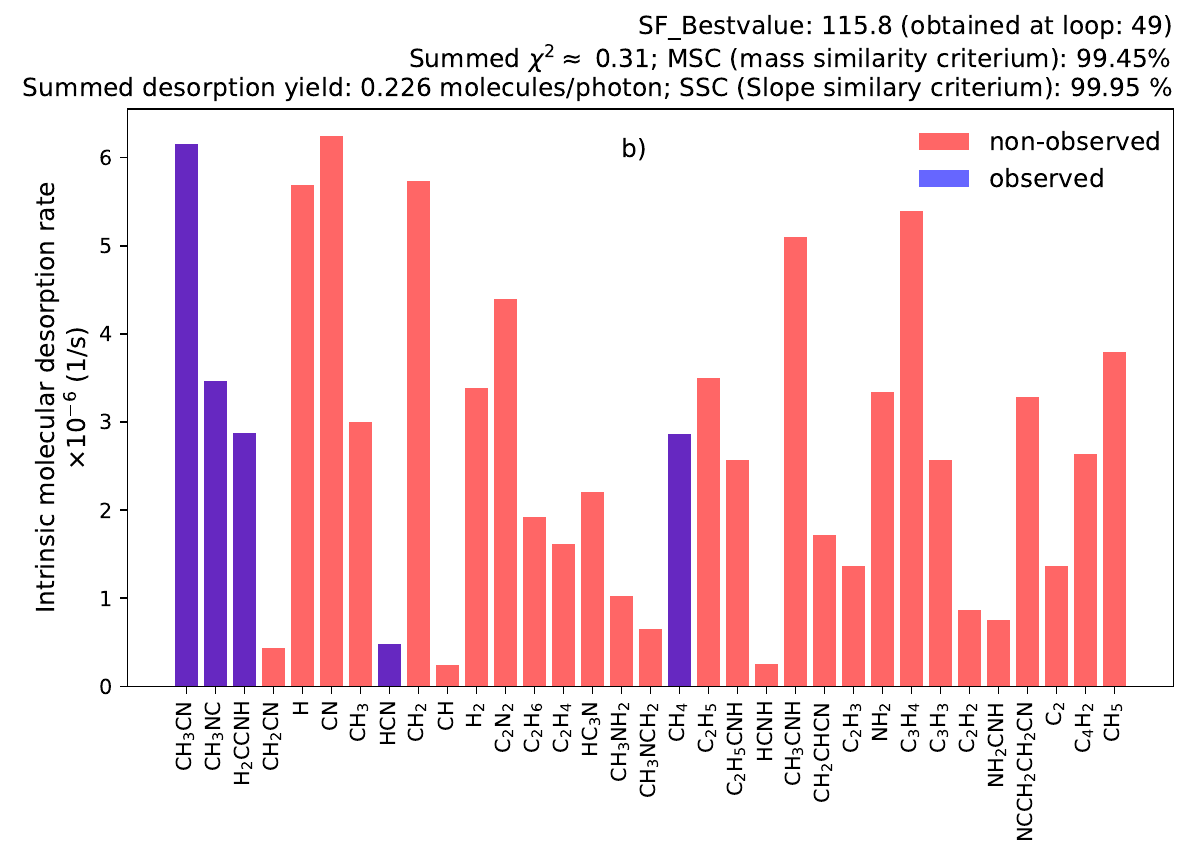}
    \caption{\label{desorption_CE} Panel a) molecular desorption of all molecular species  at chemical equilibrium stage (or equilibrium branching ratio - EBR). Panel b) intrinsic molecular desorption rates at EBR. In both panels, observed species are depicted in blue, while non-observed species are showed in red.}
\end{figure*}

\section{Results}\label{sec3}

In figure \ref{concentration_zooms}, we recall the best fit model obtained using the PROCODA code for the acetonitrile X-ray irradiation data. The figure shows the evolution of chemical concentrations for the 33 molecular species considered in the chemical reaction network. It is easy to see, particularly in panels b and c, that at some points the curves cross each other, indicating that one species becomes more abundant than the other. This is important when analyzing the desorption column densities, see Eq. \eqref{des_eq}. 

In figure \ref{desorption_density}, the desorbed column densities for the 33 modeled species are presented. The desorbed column density of acetonitrile is at least two orders of magnitude larger than any other one. This indicates that the desorption processes are dominated by the acetonitrile molecule. However, from panel a of figure \ref{concentration_zooms} it is also noticeable that acetonitrile concentration remains larger than the other ones during the entire irradiation procedure. This is understandable, once we have modeled the desorbed column densities as proportional to the concentrations, see also figure \ref{concentration_zooms}. 

By comparing the CH and CH$_3$ concentrations, we can see that the difference in desorption rates can sometimes overcome the difference in concentrations, so a molecule with a smaller concentration can have a larger desorbed column density. The desorption rate for CH$_3$ is approximately 20 times larger than that for CH species. This difference in desorption rates is the main factor responsible for the higher desorbed column density for CH$_3$. So, it is noticeable that larger concentrations of a given species does not imply a larger desorption column density, this can also be perceived by comparing the results for H and CN species.

\begin{table}[H]
\caption{\label{desorption-rates} Intrinsic molecular desorption rates for each molecular species considered in the chemical reaction network.} 
\begin{tabular}{|p{4cm}|p{3cm}|}
\hline
Molecular species & Intrinsic Desorption Rate $k_{\rm DES,i}$ ($10^{-6}$ s$^{-1}$)\\
\hline
CH$_3$CN &  $6.2$\\
\hline
CH$_3$NC & 3.5\\
\hline
H$_2$CCNH & 2.9\\
\hline
CH$_2$CN & 0.4\\
\hline
H & 5.7\\
\hline
CN & 6.2\\
\hline
CH$_3$ & 3.0\\
\hline
HCN & 0.5\\
\hline
CH$_2$ & 5.7\\
\hline
CH & 0.2\\
\hline
H$_2$ & 3.4\\
\hline
C$_2$N$_2$ & 4.4 \\
\hline
C$_2$H$_6$ & 1.9\\
\hline
C$_2$H$_4$ & 1.6\\
\hline
HC$_3$N & 2.2\\
\hline
CH$_3$NH$_2$ & 1.0\\
\hline
CH$_3$NCH$_2$ & 0.7\\
\hline
CH$_4$ & 2.9\\
\hline
C$_2$H$_5$ & 3.5\\
\hline
C$_2$H$_5$CNH & 2.6\\
\hline
HCNH & 0.3\\
\hline
CH$_3$CNH & 5.1\\
\hline
CH$_2$CHCN & 1.7\\
\hline
C$_2$H$_3$ & 1.4\\
\hline
NH$_2$ & 3.3\\
\hline
C$_3$H$_4$ & 5.4\\
\hline
C$_3$H$_3$ & 2.6\\
\hline
C$_2$H$_2$ & 0.9\\
\hline
NH$_2$CNH & 0.8\\
\hline
NCCH$_2$CH$_2$CN & 3.3\\
\hline
C$_2$ & 1.4\\
\hline
C$_4$H$_2$ & 2.6\\
\hline
CH$_5$ & 3.8\\
\hline
\end{tabular}
\end{table}

The H and CN concentration curves cross each other at around 6700 s after the beginning of the experiment, but desorption column density curves cross each other only at the instant 10500 s, which represents a 3800 seconds delay. The desorbed column density of a given species depends on its concentration, desorption rate and surface coverage, see \eqref{des_eq}. The surface coverage of a given molecular species is the ratio between its concentration to the sum of all molecular concentrations. So, if the desorbed column density of CN is higher than the H one, even when the concentration of H becomes higher, that can be addressed to the desorption rates. In fact, the intrinsc desorption rates for the species are presented in Table \ref{desorption-rates}. From the table, one can note that the intrinsic desorption rates for H and CN species are slightly different, that is the reason why the desorption column density of H takes a certain time to overcome the CN one. It is also important to note that the H species takes a longer time to reach a steady state compared to the CN species, so the H species surpass the CN one in concentration and in desorption column density. If experimental data for H and CN were given, these results would be more conclusive. 

The best-fit model for the experimental data employing X-rays gives also a desorption yield, often called sputtering yield. This value was determined experimentally by employing an associative exponential equation to describe the evolution of molecular abundances of observed molecules \cite{Carvalho2020Sep}. With the help of mass conservation principle and assuming desorption column mass to be related only to the desorption of acetonitrile, $M_{\rm DES}=\Bar{MM_{\rm CH_3CN}}YF$, where $\Bar{MM_{\rm CH_3CN}}$ is the average molecular mass of acetonitrile, $Y$ is the desorption yield and $F$ is the photon fluence, \cite{Carvalho2020Sep} obtained a total desorption yield around 0.29 molecule per photon. The value for desorption yield obtained within the best-fit model was around 0.2 with an estimated $\pm$20\% error, which is in good agreement with the experimental one.

Once the abundance of the father molecule acetonitrile is larger than the other ones, the acetonitrile desorbed column density dominates the total desorbed column density. One can observe that the desorbed column density of a given species is highly related to the concentration or abundance of that species. It is noticeable that the best-fit model predicts H, CH, CN, and CH$_2$ as between the major photo-products of acetonitrile irradiation, while from infrared spectroscopy only HCN, CH$_4$ and acetonitrile isomers were identified. The H atom is not detectable from infrared spectroscopy, whereas CH and CH$_2$ radicals have no reference spectra to be identified. However, the CN radical has an infrared band close to 2085 cm$^{-1}$, which was used to identify HCN. The computational results presented here and the heating of acetonitrile sample \cite{Carvalho2022Feb}, that revealed the appearance of more infrared bands related to CN radical, indicates that 2085 cm$^{-1}$ band is probably blended. This may introduce some error in our calculations. 

In figure \ref{desorption_CE}, the molecular desorption at chemical equilibrium for each species is given in panel a, whereas in panel b the intrinsic desorption rates are listed. The intrinsic desorption rates are determined by $k_{\rm DES,i}$, given in equation \eqref{des_eq}. The larger intrinsic desorption rates are related to the species CH$_3$CN, H, CN and CH$_2$. From panel a of figure \ref{desorption_CE}, the molecular desorption at chemical equilibrium for acetonitrile represents almost 98\% of the total, and the other ones are much smaller with emphasis to H, CN and CH$_2$ species that virtually accounts for the other $\sim 2$\%. In table \ref{desorption-yields}, we compare the desorption yields of this work with the ones presented by \cite{Basalgete2023Aug} for some species. In particular, the desorption yield for acetonitrile obtained here is almost 3 orders of magnitude higher, however for HCN and CH$_3$ is between 1 or 2 orders smaller. The main difference between the experimental setups is the beam line. Authors of \cite{Basalgete2023Aug} used a 420 eV beam line, while in \cite{Carvalho2020Sep} authors used a broadband beam line ranging from vacuum ultraviolet to soft X-rays (6 eV - 2 keV), see figure 2 of \cite{Pilling2015Sep}. The broadband feature can be responsible for the higher desorption yield of acetonitrile because the large range of vacuum ultraviolet photons can deposit more energy on the surface, thus increasing the photodesorption process. The large energy range can also have inner-shell resonance photons (C1s at 290 eV and N1s at 400eV). Such resonance photons increase the interaction cross-sections with the radiation and also could allow greater desorption of acetonitrile molecules.

\begin{table*}
\caption{\label{desorption-yields} Desorption yields for the irradiation of pure acetonitrile ices with X-rays.} 
\begin{tabular}{|p{2cm}|p{2.5cm}||p{3cm}||p{5cm}|}
\hline
Molecular species & Desorption yield (molecules/photon) & X-ray beam line & References\\
\hline
CH$_3$CN &  $0.2\pm 0.04$ & $6-2000$ eV & \cite{Carvalho2020Oct} and this work\\
 & $5.2\pm 1.5\times 10^{-4}$ & 420 eV & \cite{Basalgete2023Aug}\\
\hline
HCN & $5.0\pm 1.0\times 10^{-6}$ & $6-2000$ eV & \cite{Carvalho2020Oct}\\
& $2.5\pm 0.3\times 10^{-3}$ & 420 eV & \cite{Basalgete2023Aug}\\
\hline
CH$_3$ & $4.0\pm 0.8\times 10^{-5}$ & $6-2000$ eV & \cite{Carvalho2020Oct}\\
& $1.3\pm 0.7\times 10^{-3}$ & 420 eV & \cite{Basalgete2023Aug}\\
\hline
\end{tabular}
\end{table*}

\section{Astrophysical implications}
\label{sec4}
\begin{figure*}
  \centering
  \includegraphics[width=0.8\linewidth]{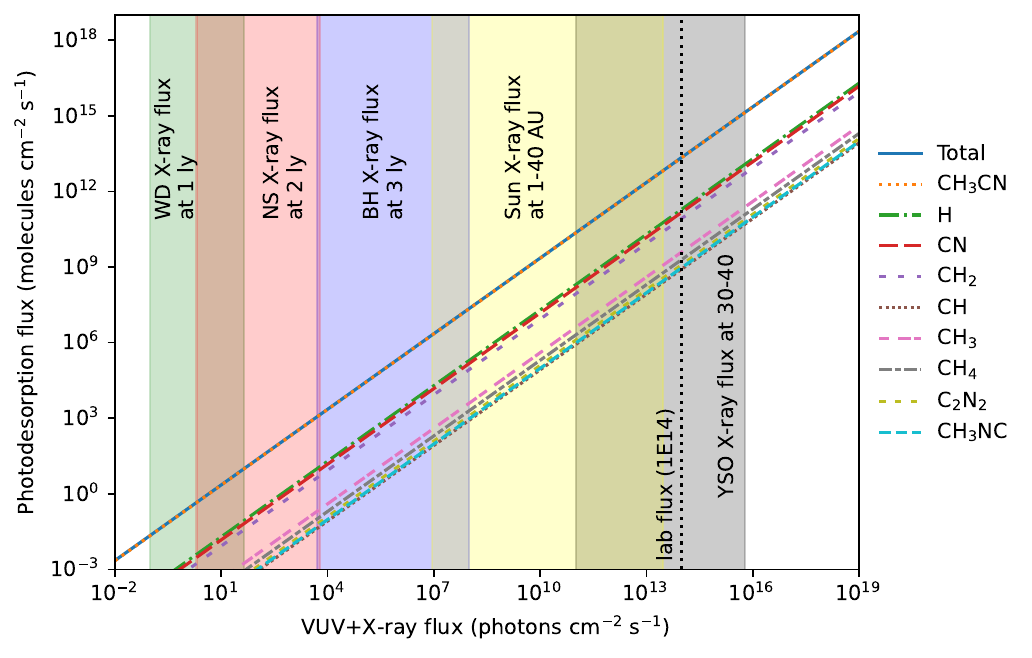}
    \caption{\label{desorption_flux} Desorption flux as a function of photon flux in the range of 6eV$-$2keV for selected molecular species with higher abundances at chemical equilibrium and high desorption rates. The vertical dotted line marks the broadband X-ray flux of 10$^{14}$ photons cm$^{-2}$ s$^{-1}$, which corresponds to laboratory photon flux used during the ice irradiation experiment. The color filled regions correspond to: typical white dwarf X-ray fluxes at 1 ly distances (green), typical neutron star X-ray fluxes at 2 ly distances (red), typical black hole X-ray fluxes at 3 ly distances (blue), Sun's X-ray fluxes from 1 AU to 40 AU (yellow) and YSO models of X-ray flux at $\sim$ $30-40$ AU (gray), see text for more details.}
\end{figure*}

We have seen that the amount of molecules desorbed to gas phase strongly depends on the concentration of that species, but also on the desorption rate and the time to reach chemical equilibrium. So, the desorption yield for each volatile species will depend on the radiation field, and the results presented here can be extrapolated to astrophysical environments, such as molecular clouds and protoplanetary disks. The results of table \ref{desorption-rates} are also useful to improve astrochemical modeling that sometimes is based on gas phase simulations instead of ice phase.

\begin{figure*}
  \centering
  \includegraphics[width=0.99\linewidth]{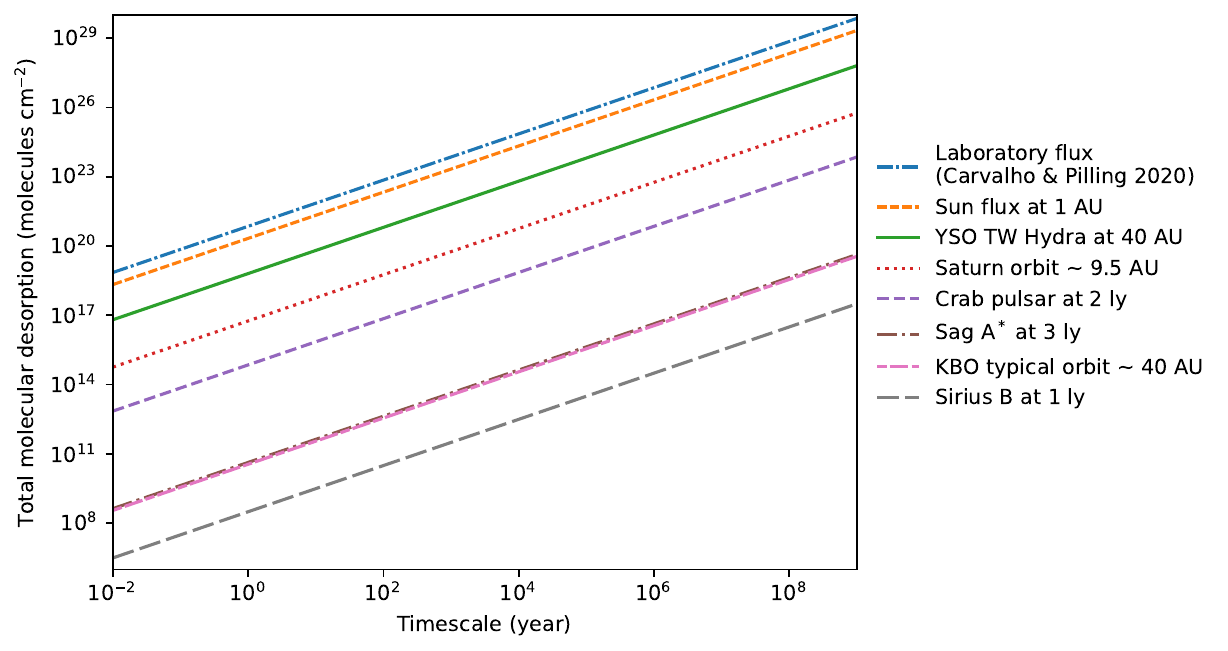}
    \caption{\label{desorption_env} Total molecular desorption induced on acetonitrile-rich ices by broadband X-ray bombardment as a function of timescale after reaching chemical equilibrium, see Eq \eqref{des_eq}. Each curve correspond to a given fixed photon flux related to several astrophysical objects.}
\end{figure*}

From table \ref{desorption-rates}, one can see that the largest intrinsic desorption rates are: CH$_3$CN ($6.2\times 10^{-6}$ s$^{-1}$), CN ($6.2\times 10^{-6}$ s$^{-1}$), CH$_2$ ($5.7\times 10^{-6}$ s$^{-1}$), H ($5.7\times 10^{-6}$ s$^{-1}$), C$_3$H$_4$ ($5.4\times 10^{-6}$ s$^{-1}$), CH$_3$CNH ($5.1\times 10^{-6}$ s$^{-1}$) and C$_2$N$_2$ ($4.4\times 10^{-6}$ s$^{-1}$). Hence, from the results of figure \ref{desorption_density} and table \ref{desorption-rates}, it is expected that gas phase observations of acetonitrile to be related especially to the species H, CN, CH$_2$, and C$_2$N$_2$, so astrophysical observations could search for them. In fact, some of these molecules are often associated or concomitantly detected in astrophysical environments together with acetonitrile, see e.g. \cite{Woodney2002May}. The adenine nitrogenous basis has been shown to survive solar X-ray fluxes during dozens of thousand of years \cite{Pilling2011Mar}. Here, the photodesorption of X-ray irradiated acetonitrile ice leads to gas phase H, CN and C$_2$N$_2$, which can be combined to form adenine molecules via pentamerization processes \cite{Roy2007Oct}.

The photodesorption flux, $\Phi$, of all molecular species after reaching chemical equilibrium can be estimated as
\begin{equation}
    \Phi(F)= YF,
\end{equation}
where $F$ means photon fluence and $Y$ represents the sputtering yield obtained by the best fit model ($Y=0.2\pm 0.04$ molecules per photon). Considering the photon flux, $\phi$, to be constant, one obtains
\begin{equation}\label{des}
    \Phi(t)= Y\phi t,
\end{equation}
where $t$ represents time passed after chemical equilibrium. This gives an estimate of the timescale for photodesorption processes to occur in astrophysical environments. The total desorption yield can be fragmented into each desorption yield for individual species, to do this we multiply the total desorption yield, $Y$, by the desorption fraction of a given species, i.e., ${\rm DES}_i/\sum_i {\rm DES}_i$, so, the desorption yield for that species will be given by the total desorption yield multiplied by its respective fraction. Finally, the photodesorption flux for a given species is calculated using equation \eqref{des} with its respective desorption yield.

Figure \ref{desorption_flux} presents the photodesorption flux, individual and total, as a function of the photon flux. The colored filled regions mark different domains of X-ray fluxes, related to the snow region of several objects, from compact to ordinary and young stars. The dotted vertical line marks the broadband X-ray flux used during ice irradiation experiment \cite{Carvalho2020Oct}. From the figure, one can see that the desorption flux induced by X-ray radiation field of white dwarfs at its snowline is in the order of $10^{-3}-10^{0}$ molecules cm$^{-2}$ s$^{-1}$, while the desorption flux for neutron star snowline is in the order of $10^{-1}-10^3$ and black holes desorption flux are between $10^3-10^7$, and the photodesorption flux for the Sun and YSO objects are between $10^6-10^{15}$ molecules cm$^{-2}$ s$^{-1}$. In particular, at a distance of 9.5 AU (Saturn orbit) the desorption flux is $1.6\times 10^9$ molecules cm$^{-2}$ s$^{-1}$, which helps to understand desorption processes induced in Enceladus moon.

In figure \ref{desorption_env} is presented the total column density desorbed as a function of timescale after chemical equilibrium. Some fixed values for photon flux were considered, which are related to the X-ray flux astrophysical ices would be subject, e.g., the X-ray flux produced by the Sun at 1 AU, 9.5 AU (Saturn orbit) and 40 AU (KBOs typical orbit), and the X-ray flux of some compact objects at snowline distances. From the figure, it is noticeable that molecular desorption spans over a wide interval ($10^8-10^{20}$ molecules per cm$^2$) for the one-year timescale. At Saturn's orbit, the one-year timescale corresponds to a total desorption of $10^{16}$ molecules per cm$^2$. These results are similar to the ones presented in \cite{Andrade2010Dec,Almeida2014Mar,Pilling2019Sep}.

The Horsehead Nebula presented CH$_3$CN spectral lines that are 40 times brighter than those in the dense core, suggesting that thermal processes are not enough to explain these abundances \cite{Gratier2013Sep}. The VUV photodesorption may explain the high gas phase abundance of acetonitrile in the Horsehead Nebula, according to astrochemical modeling \cite{LeGal2017Sep}. The photodesorption yields obtained here are higher than the ones presented in \cite{Basalgete2021Dec,Basalgete2023Aug} and contribute to the understanding that desorption explains the gas phase abundances of acetonitrile in the Horsehead Nebula. The desorption rates obtained here can also be used to improve the previous results, while also helping to interpret gas phase acetonitrile observations expected to be made by James Webb Satellite Telescope.

\section{Conclusions}
\label{sec5}
From a previously used chemical reaction network for acetonitrile irradiated ice, we derive the column densities of observed and non-observed molecular species during the experiment and also the rate constants for the chemical reactions. Here, as a compliment, we derive the desorbed column densities of observed and non-observed species and give also the desorption rates for each. These results can help to improve astrochemical modeling, and are relevant for analyzing astrophysical scenarios, such as the surface of Saturn moon, Enceladus, which has nitrogen and oxygen compounds that are precursors of amino-acids, the building blocks of life as we know.

The desorption process for a given species depends strongly on its concentration. In that case, after X-ray irradiation of a pure acetonitrile ice, the concentrations of all species are small compared to the acetonitrile one, so essentially the desorbed column densities of other species are also small. However, although acetonitrile dominates the total desorbed column density, the CN molecule has an intrinsic molecular desorption rate similar to the acetonitrile one. Other molecules with particular high intrinsic molecular desorption rates were: H, H$_2$, CH$_2$, C$_2$N$_2$, CH$_3$CNH and C$_3$H$_4$.

The higher values of intrinsic desorption rates of CN and C$_2$N$_2$ helps to explain the infrared absorbance band of 4.8 $\mu$m as HCN. At experimental chemical equilibrium, the acetonitrile molecular desorption is still much larger than any other one. Excepting for acetonitrile, the molecules with higher molecular desorption at chemical equilibrium are H, CN and CH$_3$.

The desorption flux and the timescales of molecular desorption were derived for several astrophysical environments, such as the Saturn's orbit, where the desorption flux will be between $10^6-10^9$ molecules cm$^{-2}$ s$^{-1}$ and the molecular desorption for a one-year timescale is around 10$^{16}$, results that are relevant for Enceladus. The results obtained here helps also to explain the gas phase abundance of acetonitrile in the Horsehead Nebula.

\section*{Acknowledgements}

The authors acknowledge the financial support of Funda\c{c}\~ao Arauc\'aria under the project Novo Arranjo de Pesquisa e Inovação-NAPI ``Fen\^omenos Extremos no Universo''. The authors acknowledge also the Brazilian research agencies Conselho Nacional de Desenvolvimento Científico e Tecnológico—CNPq (\#306145/2015-4; 302985/2018-2) and Coordenação
de Aperfeiçoamento de Pessoal de Nível Superior—CAPES (\#PNPD/88887.368365/2019-00; \#PNPD/88887.751149/
2022-00). SG acknowledges Funda\c{c}\~ao Arauc\'aria for financial support.

%%%%%%%%%%%%%%%%%%%%%%%%%%%%%%%%%%%%%%%%%%%%%%%%%%
\section*{Data Availability}

The data that support the findings of this study are available from the corresponding author, GAC, upon reasonable request.



\bibliographystyle{mnras}
\bibliography{example} % if your bibtex file is called example.bib

%\begin{thebibliography}{99}
%\end{thebibliography}

%%%%%%%%%%%%%%%%%%%%%%%%%%%%%%%%%%%%%%%%%%%%%%%%%%

%%%%%%%%%%%%%%%%% APPENDICES %%%%%%%%%%%%%%%%%%%%%

%\appendix



%%%%%%%%%%%%%%%%%%%%%%%%%%%%%%%%%%%%%%%%%%%%%%%%%%

% Don't change these lines
\bsp	% typesetting comment
\label{lastpage}
\end{document}